\documentclass[12pt,preprint]{aastex}

\shorttitle{Magnetic Fields in Dark Cloud Cores}
\shortauthors{Troland \& Crutcher}

\begin{document}


\title{Magnetic Fields in Dark Cloud Cores: Arecibo OH Zeeman Observations}


\author{Thomas H. Troland}
\affil{Physics and Astronomy Department, University of Kentucky,
    Lexington, KY 40502}
\email{troland@pa.uky.edu}
\and
\author{Richard M. Crutcher}
\affil{Astronomy Department, University of Illinois, Urbana, IL 61801}
\email{crutcher@uiuc.edu}


\begin{abstract}
We have carried out an extensive survey of magnetic field strengths toward dark cloud cores in order to test models of star formation: ambipolar-diffusion driven or turbulence driven. The survey involved $\sim500$ hours of observing with the Arecibo telescope in order to make sensitive OH Zeeman observations toward 34 dark cloud cores. Nine new probable detections were achieved at the 2.5-sigma level; the certainty of the detections varies from solid to marginal, so we discuss each probable detection separately. However, our analysis includes all the measurements and does not depend on whether each position has a detection or just a sensitive measurement. Rather, the analysis establishes mean (or median) values over the set of observed cores for relevant astrophysical quantities. The results are that the mass-to-flux ratio is supercritical by $\sim 2$, and that the ratio of turbulent to magnetic energies is also $\sim 2$. These results are compatible with both models of star formation. However, these OH Zeeman observations do establish for the first time on a statistically sound basis the energetic importance of magnetic fields in dark cloud cores at densities of order $10^{3-4}$ cm$^{-3}$, and they lay the foundation for further observations that could provide a more definitive test.
\end{abstract}





\section{Introduction}

It has become increasingly clear that cosmic magnetic fields are pervasive, ubiquitous, and likely important in the properties and evolution of almost everything in the Universe, from planets to quasars \citep{WB05}. However, the role played by magnetic fields in star formation has remained uncertain. One extreme of the theories of star formation is that magnetic fields control the formation and evolution of the molecular clouds from which stars form, including the formation of cores and their gravitational collapse to form protostars. Detailed theoretical work has been carried out by a number of groups. \citet{Shuetal99}, \citet{McKee99}, and \citet{MC99} have reviewed and summarized the state of this theory. The fundamental principle is that clouds are formed with subcritical masses, $M < M_\Phi$, where $M_\Phi = \Phi/2\pi\sqrt{G}$, $\Phi$ is the magnetic flux, and the expression for $M_\Phi$ is from \citet{nn78}. The magnetic field is frozen only into the ionized gas and dust; neutral gas and dust contract gravitationally through the field and the ions, accumulating mass (but not flux) in the cloud cores. This process is known as ambipolar diffusion. When the core mass reaches and exceeds $M_\Phi$, the core becomes supercritical ($M > M_\Phi$), collapses, and forms stars. The magnetic flux mostly remains behind in the envelope. Because the ambipolar diffusion time scale for the formation of cores is fairly long ($\geq 10^7$ yr), molecular clouds would have long lifetimes, although observable ``starless'' core lifetimes would be short, $\sim 10^6$ yr \citep{TM04}.

This ambipolar-diffusion driven model for star formation was the standard for many years, but doubt of its validity was raised by efforts to determine molecular cloud and core lifetimes. \citet{Beichmanetal86} used the number of molecular cores detected at various densities to determine typical statistical timescales for evolutionary sequences. Then the ratio of the number of molecular cores with T Tauri stars to the number of starless cores, together with typical T Tauri star lifetimes, led to an estimate of the lifetimes of starless cores. The result was a few times $10^6$ yr. This early estimate was refined by \citet{LM99} to $\sim 0.3 - 1.6 \times 10^6$ yr. Later workers \citep{KWA05, Kandorietal05} inferred similarly short core lifetimes. This led to a new theory -- that molecular clouds are intermittent phenomena, with short ($\sim 10^6$ yr) lifetimes. In this theory clouds form at the intersection of turbulent supersonic flows in the interstellar medium. Generally, clouds do not become gravitationally bound, and they dissipate. Star formation occurs only in the small fraction of the molecular gas that is sufficiently dense to be self-gravitating \citep{E00}. Magnetic fields may be present in this theory, but they are too weak to be energetically important. The role of turbulence in the energetics of the interstellar medium has been a very active area.  \citet{ES04} have written an excellent review of interstellar turbulence, and \citet{MK04} have extensively reviewed arguments that supersonic turbulence controls star formation.

Although the cloud lifetime estimates may seem to have doomed the strong magnetic field, quasi-static picture, that is not the case. First, cloud lifetimes seem to be longer than the free-fall time by a factor 2-5. Although supersonic turbulence may in principle provide support against collapse and lengthen cloud lifetimes, simulations have shown that such supersonic turbulence will damp on roughly a free-fall time scale \citep{Macetal98, OGS99}. Something else, perhaps magnetic pressure, is slowing the collapse. Moreover, \citet{TM04} have argued that the long cloud lifetimes that are quoted for the strong magnetic field model are for the entire lifetime from molecular cloud formation at density $\sim 10^2$ cm$^{-3}$ to protostar formation, and that most of that time is spent in increasing densities to the values where cores can be identified observationally, $\sim 10^4$ cm$^{-3}$. Once dense cores form, they argue that cores are magnetically critical to slightly supercritical, and that the core lifetimes in this model are $\sim 10^6$ yr, in agreement with observations.

Hence, the evidence for cloud and core lifetimes does not provide a conclusive test. Because core lifetimes appear to be several times longer than free-fall times, there is evidence for a support mechanism in cores. However, this does not prove that magnetic fields provide that support. Moreover, the lifetime numbers are somewhat uncertain, due to the statistical nature of the arguments and the chain of reasoning from T Tauri star lifetimes to ages of cores at different densities. Finally, the core lifetime estimates do not address directly the timescale of the formation of dense cores from much lower density gas.

The direct approach to resolving this uncertainly in the process by which stars form is to measure magnetic field strengths in molecular clouds in order to see whether they are weak (supercritical) or strong (subcritical).  The crucial parameter is the ratio of the mass to the magnetic flux, $M/\Phi$, which is of course closely related to $M_\Phi$. If $M/\Phi$ is observed to be supercritical, particularly at lower densities, the magnetic support model is not viable. On the other hand, if it is observed to be subcritical, magnetic fields would be too strong for the intermittent, turbulent theory to hold. The $M/\Phi$ parameter provides in principle a direct, simple, and definitive test to discriminate between the two theories of star formation.

Of course, much work has been done to measure $M/\Phi$ in molecular clouds. \citet{cr99} summarized the data available to that time, and concluded that $M/\Phi$ was approximately critical to slightly supercritical in molecular clouds. \citet{Betal01} extended the OH Zeeman work to the southern hemisphere with the Parkes telescope and added one and perhaps two new detections; they found results for $M/\Phi$ that essentially agreed with the \citet{cr99} conclusion. \citet{Cr07} has updated the discussion. However, the extant observations have a major deficiency -- a very small number of measurements of magnetic field strengths have been made in dark cloud cores -- the sites of low-mass star formation. \citet{Goodetal89} detected the OH Zeeman effect toward B1 with the Arecibo telescope. \citet{Cru93} carried out a major and very sensitive OH Zeeman survey with the NRAO 43-m telescope toward 12 dark cloud cores, but they were only able to confirm the B1 detection and to obtain a possible detection toward $\rho$ Oph. The large $18^\prime$ beam of the 43-m telescope meant that cores were not isolated; rather, the beam was filled by the cloud envelopes surrounding the cores. The $3^\prime$ beam of the Arecibo telescope used for the survey reported here is well matched to the sizes of dark cloud cores at distances of a few 100 pc. And in the early stages of the OH Zeeman survey reported here, \citet{ct00} detected the OH Zeeman effect toward L1544. Hence, there are almost no measurements of magnetic field strengths in dark cloud cores.

In this paper we report the full results of an extensive ($\sim500$ hours of telescope time) survey of the OH Zeeman effect toward dark cloud cores with the Arecibo telescope. In \S 2 we describe how the target cores were selected and the details of the observations. In \S 3 we present the results, including the inferred column densities, line-of-sight magnetic field strengths, and mass-to-flux ratios. In \S4 we discuss these results, and in \S 5 present conclusions.

\section{Observations}

\subsection{The Zeeman Effect}

\citet{Cr07} reviewed the various techniques and results for studying  magnetic fields in molecular clouds. Of these techniques, the Zeeman effect provides the only direct method for measuring magnetic field strengths in  molecular clouds. In general, only those species with an unpaired electron will have a strong Zeeman splitting. This has limited detections to the the 21-cm line of H~I, the 18-cm, 6-cm, 5-cm, and 2-cm $\Lambda$-doublet lines of OH, and the 3-mm N=1$\rightarrow$0 lines of CN. The sole exception is the 1.3-cm H$_2$O maser line, due to very strong line strengths and strong fields in H$_2$O maser regions.

Except for some OH masers, the Zeeman splitting is a small fraction of the line width, and only the Stokes $V$ spectra can be detected \citep{Cru93}; these spectra reveal the sign (i.e., direction) and magnitude of the line-of-sight component B$_{los}$. By least-squares fitting the frequency derivative of the Stokes parameter $I(\nu)$ spectrum $dI(\nu)/d\nu$ to the observed $V(\nu)$ spectrum, B$_{los}$ may be inferred \citep{Cru93}.

\subsection{Selection of Targets}

Targets were selected from among molecular cores known to exist within the Arecibo declination range.  Most targets fall into the Galactic anti-center direction (RA $\approx$  03$^h$ -- 07$^h$); a few lie in the Galactic center direction (RA $\approx$ 19$^h$ -- 21$^h$).  To identify suitable cores, we first examined CO survey data \citep{dame01} to identify nearby molecular cloud complexes observable from Arecibo.  For each of these complexes, we then consulted detailed CO and NH$_3$ maps of molecular cores to identify specific targets.  For example, in the anti-center region, the L1457 cloud has been mapped in CO by \cite{zu90} and by \cite{ms97}.  The Perseus cloud has been surveyed for NH$_3$ cores by \cite{lmg94}.  \cite{oni96, oni98} have identified and described a complete sample of 40 cores in the TMC based upon C$^{18}$O observations.  The Mon OB1 region has been surveyed in CO by \cite{omt96}.  And the Rosette Nebula cloud was studied in CO by \cite{sch98}.

Prior to Zeeman observations, we conducted a short survey of OH line strengths toward the principal cores in each cloud complex.  We found that most cores in the Galactic center direction have relatively weak OH emission lines, owing, very likely, to the Galactic background continuum emission that is comparable to the excitation temperatures of the OH transitions.  The final list of targets consisted of molecular cores for which the OH lines were relatively strong (since sensitivity to $B_{los} \propto$ line strength); promising cores had to be deleted from the list due to lack of sufficient observing time. We did not want to spend very long times observing weak lines, for that would mean our survey would have few clouds. A sensitivity calculator such as that of \citet{tt90} allows one to estimate the observing time required to reach a given sensitivity. The actual sensitivity achieved varied due to variations in line strength, line width, and available telescope time. Most targets are in nearby molecular clouds (e.g Perseus and Taurus) for which the Arecibo beam samples a relatively small linear scale (e.g. $\approx0.2$ pc for a cloud at 200 pc).  A few targets (e.g. the Rosette Nebula cloud) are more distant, but they add diversity to the sample since they are part of massive star-forming regions.

\subsection{Observations}

\begin{figure}
\epsscale{1.0}
\plotone{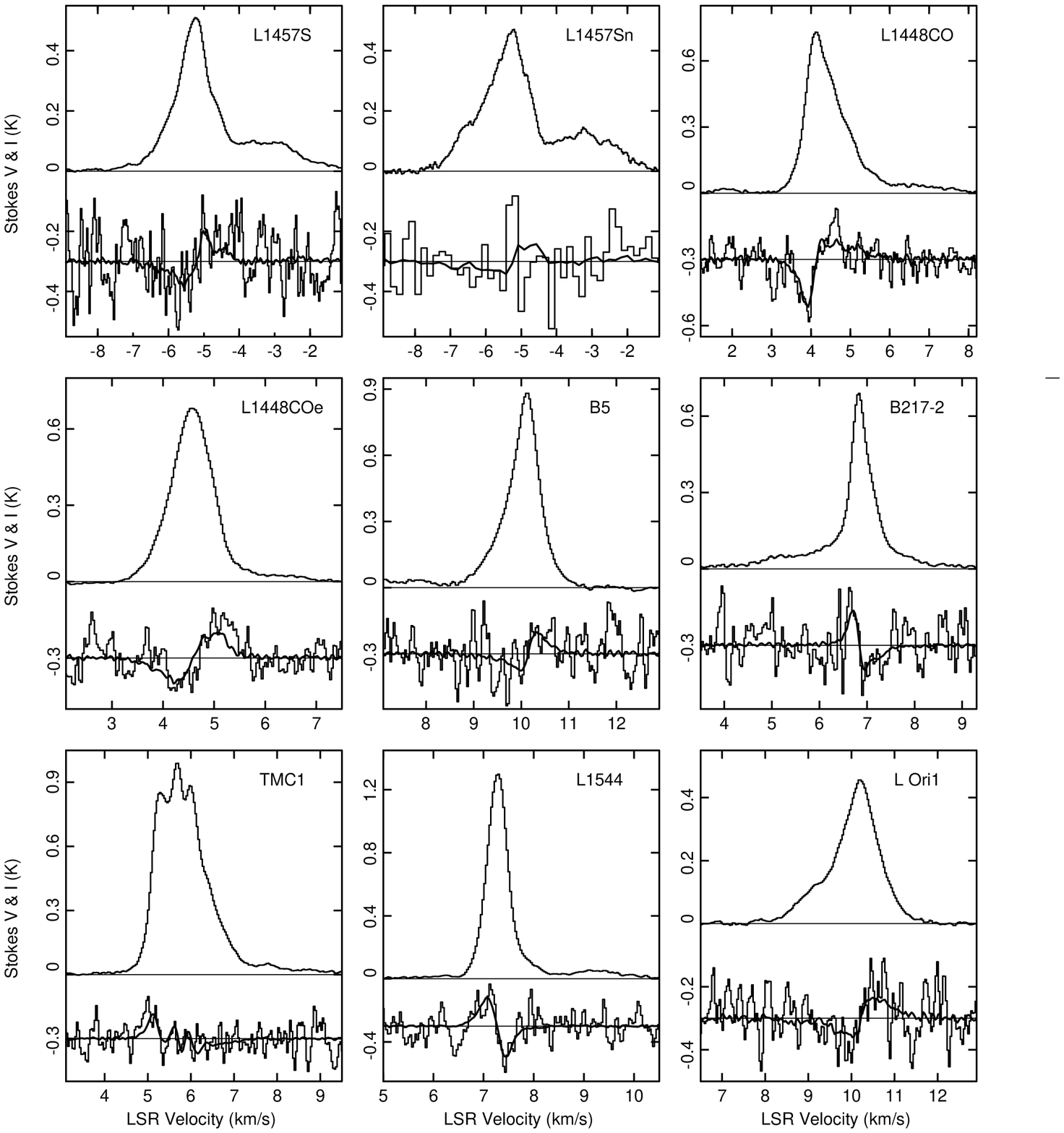}
\caption{Arecibo Stokes I and V spectra for the 9 probable detection positions in the survey. Observed data are histogram plots; fits to Stokes V are the dark lines. These are weighted means of the 1665 and 1667 results. The Stokes V spectra have been scaled up and shifted by -0.3 K for display purposes.}
\label{f1}
\end{figure}

Simultaneous observations of the 1665 and 1667 MHz OH lines were carried out in the manner described by \cite{ct00}.  Very briefly, we used the single-pixel L-band wide receiver with native linear polarizations.  A hybrid circuit immediately after the two HEMT receivers added $\pm90^\circ$ of phase shift to the two linearly-polarized outputs to convert them into orthogonal circular polarizations.  Typical system temperatures were 30-40 K.  Spectra were obtained with the ``interim'' correlator sampling 2048 spectral channels over a bandwidth of approximately 390 kHz, leading to a channel spacing of about 0.034 km s$^{-1}$.  Spectra were Hanning smoothed to produce a final spectral resolution of two channels. Data analysis followed our usual procedures for Arecibo OH Zeeman observations.  We constructed Stokes I and V profiles from the sum and difference, respectively, of the line profiles in orthogonal circular polarizations.  Then we least-squares fitted to each V profile a function defined by the sum of a constant times the I profile plus another constant times the derivative of the I profile.  The first constant is a measure of gain differences between the two circular polarizations; this constant was found to be negligible for all sources.  The second constant is proportional to the line-of-sight magnetic field component $B_{los}$.  Finally, we computed a weighted average of $B_{los}$ as derived independently from the two OH emission lines. The full survey results are given in tables~\ref{results1} and ~\ref{results2}. Figure~\ref{f1} shows the Stokes I and V spectra for the 9 positions with a Zeeman signal $> 2.5\sigma$, and figure~\ref{f2} shows all of the inferred results for B$_{los}$ from the Arecibo survey plotted against N(H$_2$). Six of the nine positions in figure~\ref{f1} have solid Zeeman detections. In order of increasing marginality of detection for the remaining probable detections, there are B217-2, B5, and L1457Sn. See detailed discussion below and in the appendix.

We carried out several tests to establish the reliability of the Arecibo system for Zeeman observations.  For one, we established the sense of circular polarization with two helical test antennas of known circular polarization sense, as described by \cite{ct00}.  We also observed the highly circularly polarized masers in W49 and compared the Stokes V profile with those in the previous literature.  Over the duration of the project, we often observed the weak but highly circularly polarized OH maser in S247 to establish that no significant change in circular polarization response had occurred.  Finally, as also described by \cite{ct00}, we estimated beam squint via continuum polarization mapping observations of an unresolved continuum source.  These data establish that the difference in pointing between the right and left circularly polarized beams (i.e. the beam squint) was of order $1^{\prime\prime}$, less than 1\% of the approximate $3^\prime$ FWHM beamwidth of the Arecibo telescope at 1666 MHz.

In addition to the Zeeman observations of molecular cores, we mapped OH line strengths in the vicinities of the Zeeman positions.  These mapping observations serve two purposes.  For one, they allow us to determine the sizes of the cores on the plane of the sky, from which core volume densities and masses may be estimated (tables 1 \& 2). Our crude OH maps were supplemented by C$^{18}$O maps from the literature. We used volume density $n(H_2) = N(H_2)/2r$ and mass $M = \pi r^2 N(H_2) 2.8 m_H$ to estimate these parameters, where r is the cloud radius and the factor 2.8 includes a 10\% He abundance and $m_H$ is the mass of an H atom. Also, the mapping observations permit us to establish upper limits for each core upon the instrumental effects of beam squint.  For this latter purpose, we used mapping positions offset approximately $3^\prime$ (one beamwidth) north, south, east and west from each Zeeman position.   From 1667 MHz profiles at these
offset positions, we estimated the magnitude and direction of the linear velocity gradient in the OH emission at the Zeeman position.  Then we computed the expected instrumental magnetic field from a beam squint of $1^{\prime\prime}$
if the position angle of the beam squint exactly matches the position angle of the velocity gradient on the sky.  We feel this procedure offers a realistic upper limit to the instrumental Zeeman effect, even if the actual beam squint is, in practice, slightly higher than $1^{\prime\prime}$.  (\citet{h99} reports beam squints of typically $1.3^{\prime\prime}$ at 1420.4 MHz; \citet{ct00} report a 1666 MHz beam squint of $1.2^{\prime\prime} \pm 0.2^{\prime\prime}$).  The actual velocity gradient is unlikely to be aligned with the telescope beam squint, especially since the beam squint position angle is nearly fixed in azimuth \citep{Hetal01}.  Therefore, the beam squint will rotate on the sky as the parallactic angle of the source changes during a Zeeman source observation.  As a result, the effective beam squint over a several hour Zeeman observation will be significantly less than the actual beam squint.

Estimates of upper limits to instrumental effects (see above) suggest that the magnetic field detections reported here are reliable.  The upper limits to instrumental effects have an average value of 2.4 $\mu$G for all of our Zeeman positions.  The highest instrumental field is 5.4 $\mu$G at position Ros4, for which $\sigma$(B) is 9.8 $\mu$G.  Of the sources listed in table 2, nine have magnetic fields with S/N $>$  2.5.  These sources are likely to be Zeeman detections on statistical grounds alone (see \S3).   In none of these sources is the upper limit to the instrumental magnetic field more than 25\% of the derived field value (table 2, column 4), for most sources it is less.  We conclude that the magnetic field detections (i.e., S/N $>$ 2.5) are quite unlikely to result from instrumental effects associated with beam squint.  Moreover, non-detections but sensitive measurements are unlikely to be significantly affected by instrumental effects. Note that \citet{Hetal01} report that beam squint is the principal contributor to instrumental circular polarization with the Arecibo L-band wide receiver.

\section{Results}

\subsection{Measurements versus Detections of $B_{los}$}

Although the astrophysical analysis we will report on below uses each of the 34 {\em measurements} of $B_{los}$ without directly considering whether each measurement is a detection or not, it is of interest to consider which of the 34 measurements should be regarded as detections. Since detections are not produced by instrumental effects (\S2), there are two statistical criteria and a third subjective criterion: (1) that $ | B_{los}|/\sigma_{B_{los}}$ be greater than an appropriate number, (2) that the 1665 and 1667 MHz lines yield the same result for $B_{los}$ within the measurement uncertainties, and (3) that a plot of Stokes V looks consistent with detection of the Zeeman effect.

For criterion (1), we set $|B_{los}|/\sigma_{B_{los}} > 2.5$, the level at which there are probably no false detections for a sample size of 34. The normal probability function says that for a normal error distribution, the fraction of the measurements that should be $2.5 \sigma$ or more from the ``real'' value is 0.0124. In our case, we have 34 positions at which we measured $B_{los}$; $0.0124 \times 34 = 0.4$. So a criterion of $2.5 \sigma$ with a sample of 34 means we would claim 0.4 false detections.  Based on this criterion alone, there are 9 detections; the probability is that there are no false detections, but it is possible (less than 50\% probability) that there is 1.

For criterion (2), we set $\Delta B / \sigma_{\Delta B} < 1.9$, where $\Delta B = | B_{los}(1665) - B_{los}(1667) |$. The normal probability function says that the fraction of the $\Delta B$ that should be $1.9 \sigma$ or more from zero is 0.057, and with 9 possible detections, this criterion would yield $0.057 \times 9 = 0.5$ possible false positives. The largest $\Delta B / \sigma_{\Delta B}$ in table~\ref{results2} is 1.7. Hence, all 9 possible detections meet this consistency criterion.

Finally, there is the subjective criterion (3). Figure~\ref{f1} shows Stokes I and V plots of each of the 9 probable detections. For these plots we have combined the 1665 and 1667 MHz results, weighted by the inverse square uncertainties in $B_{los}$ for each transition. The Stokes I spectra are the weighted sum of $T_A(1665)$ and $T_A(1667)$; these are then the Stokes I spectra that would have been observed if there were a single OH line rather than two. The Stokes V spectra are the observed Stokes V spectra for such a single line. This combination gives the appropriate spectra for judging the significance of each probable detection.

Although 9 positions pass our criteria for detection, clearly some are more solid than others. We discuss each of the 9 possible detections in the appendix with respect to the likelihood that each is or is not a detection. Again, however, note that our analysis does not depend on whether a particular position does or does not have a detected Zeeman signal.

\subsection{Estimating $B_{total}$ and $M/\Phi$ from $B_{los}$}

Following \citet{cr99}, we define
\begin{equation}
\lambda \equiv (M/\Phi)_{observed}/(M/\Phi)_{critical},
\label{lambda_def}
\end{equation}
where $(M/\Phi)_{observed}$ is the observed mass to flux ratio inferred from the ratio of $N(H_2)$ to $B$ and $(M/\Phi)_{critical}$ is the theoretically determined critical mass to flux ratio \citep{nn78}.

Then
\begin{equation}
\lambda = 7.6 \times 10^{-21} N(H_2)/B_{los},
\label{lambda_obs}
\end{equation}
where $N(H_2)$ is the column density of $H_2$ in cm$^{-2}$ and $B_{los}$ is the line-of-sight magnetic field strength in $\mu$G.

It must be kept in mind that all of the B$_{los}$ results are {\it lower} limits to the total magnetic field strength. It is possible to correct statistically for the fact that only one component of \textbf{B} is measured, i.e., B$_{{los}} = \vert{\bf B}\vert \cos\theta$. For a large number of clouds with the same total field strength for which the angle $\theta$ between {\bf B} and the observed line of sight is randomly distributed,
\begin{equation}
\overline{B}_{{los}} = \frac{\int_{0}^{\pi/2} \vert{\bf B}\vert \cos \theta \sin \theta d\theta}{\int_{0}^{\pi/2} \sin \theta d\theta} = \frac{1}{2}\vert{\bf B}\vert.
\label{Bavg}
\end{equation}

If {\bf B} is strong, clouds will have a disk morphology with \textbf{B} along the minor axis (cf, \citet{MC99}). To properly measure $\lambda$, one needs $B$ and $N$ along a flux tube, i.e., parallel to the minor axis. Then, as noted by \citet{cr99}, the path length through a disk will be too long by $1/\cos \theta$ and $N$ will be overestimated, while $\vert{\bf B}\vert$ will be underestimated by $\cos \theta$. Statistically,
\begin{equation}
\overline{M/\Phi} = \frac{\int_{0}^{\pi/2} (M/\Phi)_{obs} \cos^2 \theta \sin \theta d\theta}{\int_{0}^{\pi/2} \sin \theta d\theta} = \frac{1}{3}(M/\Phi)_{obs}.
\label{MFavgZ}
\end{equation}

\subsection{Other Physical Parameters}

We infer $N(H_2)$ toward each core from the OH data. The column density of OH is derived assuming the lines are optically thin \citep{Cr79}: $N(OH) = a \times T \times \Delta V \times 10^{14}$ cm$^{-2}$ K$^{-1}$ km s$^{-1}$, where $T$ is the peak line antenna temperature, $\Delta V$ is the full-width-at-half-maximum intensity of the line, $a = 8.49$ for the 1665 MHz line and 4.71 for the 1667 MHz line. The coefficients include the beam efficiency of the telescope, $\eta_B \approx 0.5$. The $N(OH)$ in table 1 are the average results from the two lines. Then $N(H_2) = N(OH)/8 \times 10^{-8}$ \citep{Cr79}.

A comparison of the total masses estimated from $N(OH)$ and $r$ (which we designate as  $M_{OH}$) with the virial masses ($M_{virial}$) may be instructive. The means for all 34 cores are $M_{OH} = 58$ $M_\odot$ and $M_{virial} = 74$ $M_\odot$; the respective medians are $M_{OH} = 16$ $M_\odot$ and $M_{virial} = 29$ $M_\odot$. Considering the mean values, the ratio $M_{OH}/M_{virial} = r N(OH)/\Delta V^2 X = 0.78$, where $X$ is the OH/H abundance ratio. The mean mass ratio would be 1 if any one of the following were true: $\overline{r}$ larger by 1.3, $\overline{N(OH)}$ larger by 1.3, $\overline{\Delta V}$ smaller by 0.88, or $\overline{OH/H} = 6 \times 10^{-8}$. Any or some combination of all of these is possible. The $r$ we use may not be the relevant $r$, both because of the way OH samples the gas and the geometrical assumption (\S2.3) made in going from $N$ and $r$ to $M_{OH}$. $N(OH)$ could be larger if the lines were slightly saturated rather than being optically thin as assumed. The relevant $\Delta V$ for the virial mass may be different from that given by OH, again because of how OH samples the H$_2$ gas distribution. And the value of OH/H may be different from the \citet{Cr79} result. Moreover, the virial mass calculation assumes virial equilibrium between gravity and kinetic motions, which may not be correct. The cores may not be in equilibrium, and the support provided by the magnetic field is not included. However, rather than attempting to discuss these points, we prefer to emphasize that the agreement in the two methods of determining mass is quite good; our estimates for astrophysical quantities such as the mass-to-flux ratio and the Alfv\`{e}nic Mach number are therefore unlikely to be dominated by systematic errors in column and volume densities.

Because we can only measure one component of the magnetic vector {\bf B}, it will be necessary to consider the mean or median values of the mass-to-flux ratio, ratio of turbulent-to-magnetic energy, and the Alfv\`{e}nic Mach number for the ensemble of dark cloud cores we observed. The values for the parameters that go into calculating these quantities are the following mean and median values, respectively: $\Delta v = 0.71, 0.71$ km s$^{-1}$, $B_{los} = 8.2, 5.8$ $\mu$G, $r = 0.31, 0.29$ pc, $N_{21}(H_2) = 4.5, 4.0$ cm$^{-2}$, and $n(H_2) = 3200, 1800$ cm$^{-3}$. These values will be used in the discussion below.

\section{Discussion}

An important motivation for this Zeeman effect survey was to better define the key processes responsible for star formation, in particular, the role of the magnetic field.  The role of the magnetic field in star formation, in turn, depends upon the ratio of magnetic energy in star forming (i.e. self-gravitating) clouds to other relevant energies.  These energies are the gravitational energy of the cloud and the energy of internal motions.  Since star forming clouds are, in general, not rotating nor collapsing, the energy of internal motions is presumed to reside in macroscopic motions usually referred to as turbulence and measured by the line widths of molecular spectral lines.   The parameter $\lambda$, described in \S3.2, is a measure of the ratio of gravitational to magnetic energies in the cloud.  The ratio of turbulent to magnetic energies is given by $\beta_{turb} = 4\pi F\rho \sigma^2/B^2$, where $\rho$ is the density, $\sigma$ is the 1-D turbulent velocity dispersion, $B$ is the total magnetic field strength, and $F$=3 for 3D turbulence. Then $\beta_{turb} = 0.32 n(H_2) [\Delta V^2 - 0.027]/3B_{los}^2$, where $n(H_2)$ is in cm$^{-3}$, $\Delta V$ is the FWHM in km s$^{-1}$, the line-of-sight field strength $B_{los}$ is in $\mu$G, and 0.027 removes the thermal contribution to the observed line widths for an assumed kinetic temperature of 10 K.

Values for these key ratios ($\lambda$ and $\beta_{turb}$) are not meaningful for individual clouds since we cannot measure the total field strength, only $B_{los}$.  It is for just this reason that a survey of many clouds is necessary to adequately characterize the role of the magnetic field in these objects.  Therefore, we use the data from the present survey to estimate mean and median values for the magnetic field strength and for $\lambda$ and $\beta_{turb}$ over the ensemble of clouds observed for this survey.

\begin{figure}
\epsscale{1.0}
\plotone{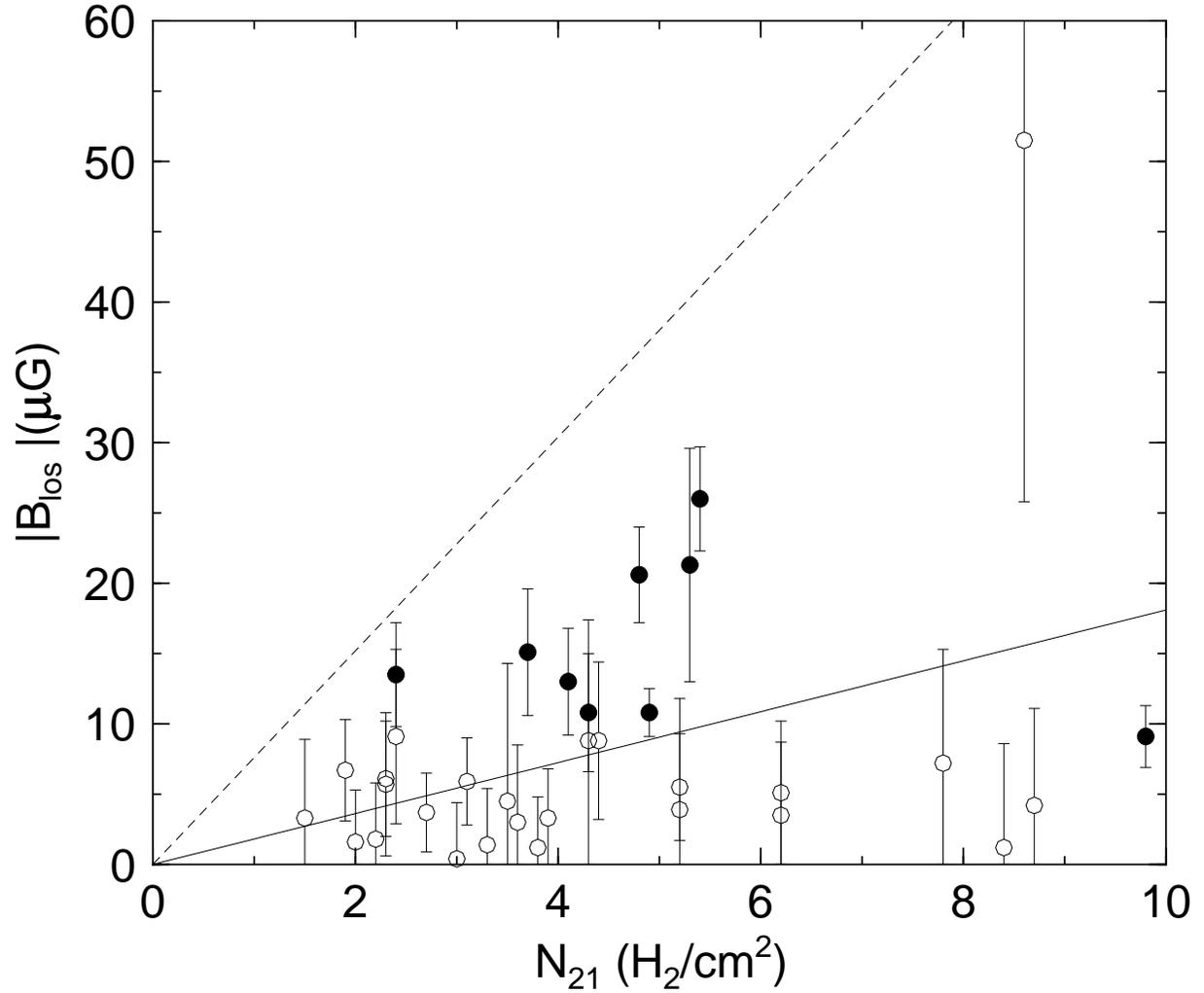}
\caption[]{Results for B$_{los}$ from the Arecibo dark cloud survey plotted against the H$_2$ column density ($N_{21} = 10^{-21} N$). The 9 probable detections (see text) are plotted as filled circles, while non-detections are plotted as open circles. Error bars are 1$\sigma$. The solid line is the weighted mean value for the mass to flux ratio with respect to critical inferred from the Zeeman $B_{los}$ data with no geometrical correction; $\lambda \approx 4.8 \pm 0.4$. After geometrical corrections (see text), $\lambda_c \approx 2$, or slightly supercritical. The dashed line is the critical mass to flux ratio.}
\label{f2}
\end{figure}

In order to proceed with the estimation of average quantities, we make the assumption that the angle between the line of sight and the magnetic field is randomly distributed among our 34 positions. For this assumption, the probability density function for $B_{los}$ is flat between 0 and the total strength $\vert{\bf B}\vert$ \citep{ht05}. That is, the observed $B_{los}$ are assumed to differ from a single total field strength for the sample of cores only by geometrical projection. We can then calculate the weighted mean $B_{los}$ and therefore $\overline{\lambda}$, where the weighting is by the inverse square uncertainties in $B_{los}$; the uncertainty in $\overline{\lambda}$ is dominated by the measurement uncertainties in $B_{los}$ rather than in $N(OH)$. The result is $\overline{B}_{los} = 8.2 \pm 2.2$ $\mu$G and $\overline{\lambda} \approx 4.2$; this is shown as a solid line in figure~\ref{f2}. If median values are used instead, we find the median value of the mass-to-flux ratio $\lambda_{1/2} \approx 5.2$. All measurements, detections and non-detections, are included in the weighted mean and median. Thus, it makes no difference for these calculations whether a given position has a detected $B_{los}$; what matters is each measured value and its uncertainty.  Our result for the total mean field strength (equation~\ref{Bavg}) is then $\vert \overline{{\bf B}}\vert = 16.4$ $\mu$G. The above value for $\overline{\lambda}$ is then systematically too large due to geometrical effects. The minimum correction is a factor of 1/2 due to the correction for only measuring $\overline{B}_{los}$. For a disk morphology, the correction is 1/3 (equation~\ref{MFavgZ}). Hence, the geometry corrected $\overline{\lambda_c} \approx 1.4 - 2.1$; $\lambda_{1/2,c} = 1.7 - 2.6$.  Finally, $\overline{\beta}_{turb} = 2.4$ and $\beta_{turb, 1/2} = 2.7$. The related Alfv\`{e}nic Mach numbers are $\overline{M}_A = 1.4$ and $M_{A,1/2} = 1.5$.

Also plotted in figure~\ref{f2} as a dotted line is the critical mass-to-flux ratio.  Although we measure only the line-of-sight component of the magnetic field, one might expect the magnetic field to point approximately along the line of sight in a few cases. If there were a subcritical core in our sample and the field pointed along the line of sight, its plotted value would lie above this critical line.

What conclusion can we draw from the estimated value of $\overline{\lambda_c}$ or $\lambda_{1/2,c}$?  The cores of molecular clouds (sampled at a typical density of a few times $10^3$ cm$^{-3}$) appear to be slightly supercritical by about a factor of 2.  Therefore, on average the gravitational energies somewhat exceed magnetic energies in these clouds, although the uncertainties in the results do not rule out a critical mass-to-flux ratio.  Had the estimated value of $\overline{\lambda_c}$ been significantly higher than 2, we could have ruled out the ambipolar diffusion model of star formation since magnetic fields, in this circumstance, would be energetically insignificant.  Likewise, had the estimated value of $\overline{\lambda_c}$ been less than unity, we could have ruled out the turbulent-driven model owing to the strong influence of magnetic fields on molecular cores.  As it is, the estimated value of $\overline{\lambda_c}$ is consistent with both extreme-case models. The ambipolar diffusion model predicts that cores are formed in subcritical clouds, but by the time cores with $n(H_2) \sim 10^4$ cm$^{-3}$ have formed by the action of ambipolar diffusion, they are critical to slightly supercritical. The turbulence driven (i.e. weak magnetic field) model forms cores by turbulent compression. Although these models have the envelopes and regions between cores highly supercritical, many cores may be only slightly supercritical. Hence, although the observations have shown that magnetic fields are sufficiently strong that they cannot be ignored, the very hard-won observational results cannot rule out either model of star formation.

What conclusion can we draw from $\overline{\beta}_{turb} = 2.4$?  Obviously, this value suggests that turbulent energy exceeds magnetic energy, just as gravitational energy exceeds magnetic energy (although not overwhelmingly in either case).  So the cloud cores sampled in this project, on average, are slightly out of equipartition between turbulent and magnetic energies.  Note that $\beta_{turb}$ is related to the Alfv\`{e}nic Mach number by the relation $M_A^2 \approx \beta_{turb}$.  Therefore, we infer a mean value $\overline{M}_A \approx 1.6$.  That is, internal motions are on average mildly super-Alfv\`{e}nic in these cores.  If turbulent energy exceeds magnetic energy by a factor of a few, (and the cores are not collapsing), then the cores are in approximate virial equilibrium between internal motions and gravitation, with magnetic support of lesser significance.  The average ratio of mass to virial mass in table 2 is 0.8, a value close to unity and consistent with this conclusion.

A concern that sometimes arises regarding Zeeman effect observations (and the conclusions drawn there from) involves tangling of the magnetic field on scales smaller than the beam.  Field tangling, of course, can reduce the beam averaged value of $B_{los}$; a field reversal within the beam can conceivably reduce the value to 0.  Although the Arecibo beam is small compared to beam sizes used for some other Zeeman effect projects, field tangling can still exist, in principle, on scales smaller than any beam.  Nonetheless, we do not believe that field tangling on scales smaller than the Arecibo beam has significantly affected this project, nor the conclusions we draw from it.  For one, higher spatial resolution studies of magnetic fields in molecular cores show little sign of field tangling.  These studies rely upon linear polarization of dust emission to map the morphology of the field in the plane of the sky.  For example, \citet{wt00} mapped linear polarization in L1544 (also included in this project), L183 and L43 at $14^{\prime\prime}$ spatial resolution.  They find a high degree of order in the field.  Likewise, \citet{grm07} find a high degree of order in the field of NGC 1333 IRAS 4A at $1.5^{\prime\prime}$ spatial resolution.  Another consideration regarding field tangling is the fact that the value of $\lambda$ is related to the net magnetic flux through a cloud.  To first approximation, at least, the beam-averaged value of $B_{los}$ is a measure of just that quantity (for a single field component), regardless of any small-scale field irregularities that may exist on scales smaller than the beam.  Finally, the data from this project suggest that the observed field strengths are sufficient to be energetically important (although not dominant) in the molecular cores.  Therefore, the observed field strengths are sufficient to impose a reasonable degree of order on the field.  If the observed field strengths are significantly lower than the actual field strengths, owing to tangling, then the actual magnetic field must be even more energetically important.  In such a case, field tangling is even less probable.  That is, the hypothesis that field strengths in molecular cores are higher than implied by this project, owing to tangling, leads to a conclusion that contradicts the hypothesis itself.

\section{Conclusions and future directions}

This project had the potential to distinguish between the two extreme models, ambipolar diffusion versus turbulence driven star formation.  However, the result for $\overline{\lambda_c}$ lies in a range compatible with both models since the molecular cores surveyed appear to be slightly magnetically supercritical.  That is, gravitational energy appears to dominate magnetic energy by a small factor.  Also, the estimated value of $\overline{\beta}_{turb} \approx 2.4$ is such that turbulent energy in the cores appears to dominate magnetic energy, again, by a rather modest factor.  To zeroth order, at least, gravitational, magnetic and turbulent energies in these cores appear to be comparable.  Once again, nature appears to have held her cards regarding magnetic effects on star formation close to her chest!  At the same time, the OH Zeeman observations reported here are of considerable value since they do establish for the first time on a statistically sound basis the energetic importance of magnetic fields in dark cloud cores at densities of order $10^{3-4}$ cm$^{-3}$.

What future observations might help to distinguish between the two extreme models of star formation? Further Arecibo observations like the ones reported here are unlikely to settle the issue, especially given the observing time that would be required to improve significantly these results.  Instead, there are at least two types of Zeeman effect observations that have the potential to do so.  One type of observation targets magnetic field strengths in the {\em envelopes} of clouds, not in their cores.  The ambipolar diffusion-driven model of star formation predicts that these envelope or inter-core regions must be magnetically subcritical ($\lambda < 1$).  The turbulence-driven model predicts that the same regions must be magnetically supercritical.  Zeeman effect measurements of OH emission lines in the inter-core regions of molecular cores would be very difficult owing to the weakness of the OH emission there.  However, OH absorption lines against extra-galactic continuum sources have the potential to probe the field along random lines of sight through the clouds.   Only the Arecibo telescope has the ability to perform this experiment since its large collecting area ensures an adequate number of background continuum sources to make the observations meaningful.  Another approach to distinguishing between the two models of star formation is to measure differential mass-to-flux ratios, that is, to measure $\lambda$ in cloud cores relative to $\lambda$ in the envelopes of the same clouds.  The ambipolar diffusion model predicts that $\lambda$ is greater in the  cores than in the envelopes, owing to the ambipolar diffusion process itself (see \S 1).  The turbulence driven model predicts the opposite, namely, that $\lambda$ decreases with increasing density \citep{dib06}.  Therefore, it should be possible to distinguish between the two models by measuring $\lambda$ in the core and envelope of the same cloud.  The OH Zeeman effect offers an opportunity to make such measurements. This approach has the advantage of eliminating geometrical effects associated with the measurement of $B_{los}$, since, to first approximation, the angle between the field and the line-of sight should be the same in both core and envelope of the same cloud.  We are currently pursuing both approaches (OH absorption and differential $\lambda$ measurements) in an attempt to better distinguish between the alternate models of star formation.

\acknowledgments
We thank the Arecibo Observatory for the generous allocation of observing time, which allowed this project to be carried out, and the observatory staff for help in making the observations successful. This research was partially supported by NSF grants AST 0205810, 0307642, and 0606822.

\appendix
\section{Discussion of possible detected Zeeman signals}

L1457S: The line has a single strong peak and an extended wing to positive velocities. The Stokes V spectrum shows that the Zeeman signal is clearly detected from the strong peak.

L1457Sn: This short integration was obtained $3^\prime$ north of the L1457N position to look for spatial variation in $B_{los}$. The line profile is qualitatively similar to that toward L1457S. Although the fitting process reported a marginally stronger field at this position than at the L1457S position, the noise level in the Stokes V data was too high to see the Zeeman signal. In order to enhance its visibility, the observed Stokes V spectrum was boxcar smoothed by 5 channels; that is the spectrum that is shown. The displayed fit line has also boxcar smoothed. Since the integration time at this position is only about 1/9th that toward L1457S, this smoothing still has a higher channel-to-channel noise level than the unsmoothed L1457S Stokes V spectrum. Yet  there is still not a clear Zeeman signal in the observed Stokes V spectrum. The fit seems to be responding to the 8 of 9 channels that are negative to the negative velocity side of the line peak and the 2 positive channels to the high velocity side. Nonetheless, we regard L1457Sn as having failed the subjective criterion (3) test. On the other hand, this position is only 1 FWHM beam away from the L1457S position; the $B_{los}$ at the two positions have the same sign (field direction) and consistent magnitude. In spite of failing criterion (3), we still regard the measurement toward L1457Sn as being more likely a detection than not, although a rather marginal one.

L1448CO: The line profile is clearly asymmetric, suggesting that there are multiple velocity components. The fit responds most strongly to the steeper slope on the lower velocity side of the line and the negative feature in Stokes V. The observed Stokes V appears to have a corresponding positive feature just to the positive velocity side of the peak, but the $dI/d\nu$ fit does not fit this feature well, due to the smaller slope of the observed Stokes I spectrum at these velocities that appears to be caused by a weaker additional velocity component at these velocities. The detected $B_{los}$ is confined to a narrow velocity component that has only its more negative side (apparently) unaffected by line component blending.

L1448COe: Observing this position, that is $3^\prime$ east of the L1448CO position, was again an attempt to look for small-scale structure in $B_{los}$. Here, unlike L1456Sn, there is a clear detection of the Zeeman signal. The Stokes I line profile shape is somewhat different from that at the L1448CO position, being at slightly more positive peak velocity and lacking the steep slope on the more negative velocity side. The close agreement in $B_{los}$ between the two positions suggests that in spite of the difference in the Stokes I profile, there is little difference in $B_{los}$.

B5: B5 has an apparently simple, single line profile. The Zeeman signal in the Stokes V spectrum is not unambiguous, but is consistent with the marginal detection given by the fit. Although this source passes our three tests, the detection should be regarded as marginal or probable rather than certain.

B217-2: The Stokes I profile is fairly simple, although there is wide, weak plateau emission that does not, however, interfere with the Zeeman signal, which appears clear and unambiguous. The less than $2\sigma$ detection of the Zeeman effect in the 1665 MHz line could lead to this being considered a marginal detection, although it fully meets our three detection criteria.

TMC1: There is clearly a complex profile, with at least three strong narrow components. Only the lowest velocity component has a clear Zeeman signal, at least on the low velocity side. The higher velocity side is confused with the middle velocity component, and there is no Zeeman signal from the higher velocity line components. This does not necessarily mean there is only a very week field in these components; blending may obscure the Stokes V Zeeman signatures from these components.

L1544: This starless core has a fairly simple line profile and an unambiguous Zeeman signal in the Stokes V spectrum.

L Ori1: This position clearly has at least two velocity components, with the unambiguous Zeeman signal coming from the stronger, more positive velocity one.

\clearpage
\begin{deluxetable}{lccccccccccc}
\tabletypesize{\footnotesize}
\tablecaption{Dark Cloud Core OH Results}
\tablewidth{0pt}
\tablehead{
\colhead{Name} & \colhead{RA\tablenotemark{a}} & \colhead{Dec\tablenotemark{b}} & \colhead{D\tablenotemark{c}} & \colhead{r\tablenotemark{d}} & \colhead{$\tau\tablenotemark{e}$} & \colhead{T$_{1665}$} & \colhead{T$_{1667}$} & \colhead{vel\tablenotemark{f}} & \colhead{$\Delta$v\tablenotemark{g}} & \colhead{N$_{21}(H_2)$\tablenotemark{h}} & \colhead{n$(H_2)$\tablenotemark{i}}
}
\startdata

L1457S &  02 56 05.9 & 19 25 05 & 300 & 0.44 & 49.3 & 0.41 & 0.65 & $-5.2$ & 1.0 & 4.1 & 1500 \\
L1457Sn &  02 56 06.0 & 19 28 05 & 300 & 0.44 & 5.7 & 0.38 & 0.60 & $-5.2$ & 1.4 & 5.3 & 2000 \\
L1448-CO &  03 25 30.5 &    30 45 43 & 300 & 0.37 & 28.1 & 0.54 & 0.99 & 4.2 & 0.93 & 5.4 & 2400 \\
L1448-COe &  03 25 44.6  &   30 45 42 & 300 & 0.37 & 31.4 & 0.52 & 0.90 & 4.6 & 0.89 & 4.8 & 2100 \\
L1455-CO &  03 27 40.1 &   30 13 03 & 300 & 0.65 & 21.2 & 0.26 & 0.53 & 4.9 & 1.5 & 4.4 & 1100 \\
N1333-8 &  03 29 02.0 &   31 13 33      & 300 & 0.46 & 8.4 & 0.30 & 0.63 & 7.8 & 1.5 & 5.2 & 1800 \\
B5 & 03 47 38.4  &   32 52 43 & 300 & 0.45 & 10.0 & 0.63 & 1.12 & 10.1 & 0.65 & 4.3 & 1600 \\
L1495(6) & 04 18 25.4 &  28 24 29 & 140 & 0.25 & 1.5 & 1.25 & 1.99 & 8.1 & 0.50 & 6.2 & 4100 \\
IRAM 04191 &  04 21 57.0  &   15 29 45 & 140 & 0.21 & 14.8 & 0.63 & 0.95 & 6.6 & 0.63 & 3.9 & 3000 \\
B217-2 &  04 28 08.6  &   26 20 53 & 140 & 0.23 & 12.8 & 0.50 & 0.86 & 6.8 & 0.47 & 2.4 & 1700 \\
L1521E &  04 32 20.0  &   26 20 25 & 140 & 0.26 & 11.1 & 0.30 & 0.56 & 6.7 & 0.71 & 2.3 & 1400 \\
L1521F &  04 28 39.8  & 26 51 35 & 140 & 0.21 & 4.9 & 0.71 & 1.16 & 6.5 & 0.46 & 3.3 & 2600 \\
L1524-2 &  04 29 31.8 & 26 59 59 & 140 & 0.40 & 8.9 & 0.72 & 1.24 & 6.4 & 0.70 & 5.2 & 2100 \\
L1524-4 &  04 30 05.7  & 24 25 16 & 140 & 0.24 & 6.3 & 0.75 & 1.30 & 6.3 & 0.49 & 3.8 & 2600 \\
L1551S2 &  04 30 57.5  & 18 15 35 & 140 & 0.18 & 4.3 & 0.63 & 1.12 & 6.6 & 0.35 & 2.3 & 2100 \\
B18-5 &  04 35 51.3 &  24 09 21 & 140 & 0.18 & 6.7 & 0.57 & 0.81 & 5.8 & 1.14 & 6.2 & 5600 \\
L1534 &  04 39 34.8 &  25 41 47 & 140 & 0.26 & 1.3 & 1.21 & 1.57 & 6.3 & 0.76 & 8.4 & 5200 \\
TMC1 &  04 41 33.0 &  25 44 44 & 140 & 0.31 & 23.7 & 0.78 & 1.24 & 5.7 & 1.26 & 9.8 & 5100 \\
L1507A1 &  04 42 38.6  & 29 43 45   & 140 & 0.26 & 6.3 & 0.77 & 1.23 & 6.2 & 0.39 & 3.0 & 1900 \\
CB23 &  04 43 31.5 &  29 39 11 & 140 & 0.21 & 11.1 & 0.53 & 0.86 & 6.1 & 0.36 & 1.9 & 1500 \\
L1544 &  05 04 16.6  & 25 10 48 & 140 & 0.12 & 15.5 & 1.04 & 1.58 & 7.2 & 0.48 & 4.9 & 6600\\
L Ori1 &  05 31 38.3  & 12 33 06 & 400 & 0.79 & 32.2 & 0.40 & 0.55 & 10.2 & 0.99 & 3.7 & 800 \\
Ros4 &  06 34 36.9  &04 12 37 & 1600 & 2.4 & 9.9 & 0.19 & 0.45 & 12.6 & 1.51 & 3.5 & 200 \\
Mon16W &  06 40 47.4 & 09 33 15 & 950 & 1.5 & 17.3 & 0.53 & 0.75 & 5.7 & 1.55 & 7.8 & 800 \\
Mon16 &  06 41 03.5 & 09 33 13 & 950 & 1.3 & 5.6 & 0.38 & 0.78 & 6.1 & 2.0 & 8.6 & 1100 \\
Mon16N &  06 41 03.6 & 09 37 13 & 950 & 1.6 & 21.5 & 0.41 & 0.89 & 5.5 & 1.82 & 8.7 & 900 \\
L723 &  19 17 53.9 & 19 12 19 & 300 & 0.46 & 7.3 & 0.34 & 0.55 & 11.0 & 1.05 & 3.6 & 1300 \\
L771 &  19 20 49.5 & 23 29 57 & 400 & 0.34 & 6.8 & 0.42 & 0.68 & 10.9 & 0.47 & 2.0 & 900 \\
L774w &  19 22 37.1 & 23 25 10 & 200 & 0.21 & 2.7 & 0.38 & 0.62 & 11.0 & 0.63 & 2.4 & 1900 \\
L774 &  19 22 51.6 & 23 25 11 & 200 & 0.21 & 10.7 & 0.49 & 0.80 & 11.0 & 0.63 & 3.1 & 2400 \\
L663 &  19 36 57.7 & 07 34 17 & 250 & 0.15 & 1.7 & 0.45 & 0.83 & 8.3 & 0.32 & 1.5 & 1700 \\
L694n &  19 41 07.1 & 10 58 08 & 250 & 0.34 & 6.7 & 0.63 & 1.00 & 9.6 & 0.43 & 2.7 & 1300 \\
L694s &  19 41 07.2 & 10 51 28 & 250 & 0.34 & 5.3 & 0.49 & 0.82 & 9.4 & 0.44 & 2.2 & 1100 \\
L810 &  19 45 24.0  & 27 51 01 & 2000 & 0.87 & 6.5 & 0.18 & 0.68 & 15.7 & 1.45 & 4.3 & 800 \\
\enddata
\tablenotetext{a}{J2000: hr min sec}
\tablenotetext{b}{J2000: $^\circ$ $^\prime$ $^{\prime\prime}$}
\tablenotetext{c}{Distance, pc}
\tablenotetext{d}{mean cloud radius to OH half-power point, pc}
\tablenotetext{e}{hours of integration time}
\tablenotetext{f}{$V_{LSR}$, km s$^{-1}$}
\tablenotetext{g}{full-width at half-maximum, km s$^{-1}$}
\tablenotetext{h}{H$_2$ column density, $H_2 \times 10^{21}$ $cm^{-2}$}
\tablenotetext{i}{H$_2$ volume density, $H_2$ $cm^{-3}$}

\label{results1}
\end{deluxetable}

\clearpage
\begin{deluxetable}{lccccccc}
\tabletypesize{\footnotesize}
\tablecaption{Dark Cloud Core Zeeman Results}
\tablewidth{0pt}
\tablehead{
\colhead{Name} & \colhead{B$_{los}$(1665)\tablenotemark{a}} & \colhead{B$_{los}$(1667)\tablenotemark{a}} & \colhead{$\overline{B}_{los}$\tablenotemark{a}} & \colhead{$\frac{|\overline{B}_{los}|}{\sigma_{B_{los}}}$} & \colhead{$\frac{\Delta B}{\sigma_{\Delta B}}$} & \colhead{$M_{OH}$\tablenotemark{b}} & \colhead{$M_{virial}$\tablenotemark{b}}
}
\startdata
L1457S & $-13.9 \pm 4.7$ & $-11.3 \pm 6.4$ & $-13.0 \pm 3.8$ & 3.4 & 0.3 & 53 & 111  \\
L1457Sn & $-12.8 \pm 10.2$ & $-37.7 \pm 14.2$ & $-21.3 \pm 8.3$ & 2.6 & 1.4 & 69 & 217  \\
L1448-CO & $-25.1 \pm 4.6$ & $-27.6 \pm 6.2$ & $-26.0 \pm 3.7$ & 7.0 & 0.3 & 50 & 81  \\
L1448-COe & $-23.4 \pm 4.3$ & $-16.4 \pm  5.3$ & $-20.6 \pm  3.4$ & 6.1 & 1.0 & 44 & 74  \\
L1455-CO & $-13.9 \pm  7.7$ & $-3.1 \pm  8.1$ & $-8.8 \pm  5.6$ & 1.6 & 1.0 & 130 & 370  \\
N1333-8 & $-0.0 \pm  10.4$ & $-9.2 \pm 12.0$ & $-4.0 \pm 7.8$ & 0.5 & 0.6 & 74 & 260  \\
B5 &  $-18.1 \pm  6.0$ & $-4.0 \pm  5.8$ & $-10.8 \pm  4.2$ & 2.6 & 1.7 & 59 & 48  \\
L1495(6) & $7.9 \pm 6.8$ & $-18.9 \pm 7.0$ & $-5.1 \pm 5.1$ & 1.0 & 2.7 & 26 & 16  \\
IRAM 04191 & $2.2 \pm  4.6$ & $4.8 \pm  5.5$ & $3.3 \pm 3.5$ & 0.9 & 0.4 & 12 & 21  \\
B217-2 & $7.0 \pm  5.4$ & $19.3 \pm  5.1$ & $13.5 \pm 3.7$ & 3.6 & 1.7 & 8.7 & 13  \\
L1521E & $-0.6 \pm  7.3$ & $11.4 \pm  7.0$ & $5.7 \pm 5.1$ & 1.1 & 1.2 & 11 & 33  \\
L1521F & $1.2 \pm  5.0$ & $-5.9 \pm  6.6$ & $-1.4 \pm  4.0$ & 0.4 & 0.9 & 9.8 & 11  \\
L1524-2 & $-11.9 \pm 5.5$ & $0.3 \pm 5.2$ & $-5.4 \pm 3.8$ & 1.4 & 1.6 & 56 & 49  \\
L1524-4 & $-0.5 \pm  4.5$ & $-2.6 \pm 6.3$ & $-1.2 \pm 3.6$ & 0.3 & 0.3 & 15 & 15  \\
L1551S2 & $9.4 \pm 5.7$ & $2.6 \pm 5.8$ & $6.1 \pm 4.1$ & 1.5 & 0.8 & 5.1 & 5.6  \\
B18-5 & $1.5 \pm 6.3$ & $-14.2 \pm 9.2$ & $-3.6 \pm 5.2$ & 0.7 & 1.4 & 13 & 59  \\
L1534 & $3.9 \pm 9.2$ &  $-3.5 \pm 12.2$ & $1.2 \pm 7.4$ & 0.2 & 0.5 & 38 & 38  \\
TMC1 & $10.7 \pm 3.0$ & $7.1 \pm 3.4$ & $9.1 \pm 2.2$ & 4.1 & 0.8 & 64 & 120  \\
L1507A1 & $7.2 \pm 6.3$ & $-5.6 \pm 5.2$ & $-0.3 \pm 4.0$ & 0.1 & 1.6 & 14 & 10  \\
CB23 & $-9.5 \pm 5.7$ & $-4.8 \pm 4.7$ & $-6.7 \pm 3.6$ & 1.9 & 0.6 & 5.7 & 6.9  \\
L1544 &  $10.8 \pm 2.4$ & $10.8 \pm 2.6$ & $10.8 \pm 1.7$ & 6.4 & 0.0 & 4.7 & 7.0  \\
L Ori1 & $-18.1 \pm 5.4$ & $-8.3 \pm 8.1$ & $-15.0 \pm 4.5$ & 3.3 & 1.0 & 160 & 200  \\
Ros4 & $-7.9 \pm 10.9$ & $10.8 \pm 23$ & $-4.4 \pm 9.8$ & 0.4 & 0.7 & 1400 & 1400  \\
Mon16W & $15.2 \pm 14.1$ & $3.2 \pm 9.9$ & $7.2 \pm 8.1$ & 0.9 & 0.7 & 1200 & 900  \\
Mon16 & $17.8 \pm 41$ & $73.4 \pm 33$ & $51.3 \pm 26.0$ & 2.0 & 1.1 & 1000 & 1300  \\
Mon16N & $-1.6 \pm 9.8$ & $9.8 \pm 9.7$ & $4.2 \pm 6.9$ & 0.6 & 0.8 & 1500 & 1300  \\
L723 & $11.9 \pm 7.6$ & $-7.1 \pm 8.1$ & $3.0 \pm  5.5$ & 0.5 & 1.7 & 51 & 130  \\
L771 & $-0.9 \pm 5.0$ & $-2.4 \pm 5.5$ & $-1.6 \pm 3.7$ & 0.4 & 0.2 & 16 & 19  \\
L774w & $-7.0 \pm 7.9$ & $-12.3 \pm 9.9$ & $-9.1 \pm 6.2$ & 1.5 & 0.4 &7.2 & 21  \\
L774 & $-2.2 \pm 4.1$ & $-10.6 \pm 4.6$ & $-5.9 \pm 3.1$ & 1.9 & 1.4 & 9.3 & 21  \\
L663 & $10.6 \pm 7.4$ & $-6.0 \pm 8.4$ & $3.4 \pm 5.5$ & 0.6 & 1.5 & 2.3 & 3.9  \\
L694n & $5.0 \pm 3.5$ & $1.3 \pm 4.7$ & $3.7 \pm 2.8$ & 1.3 & 0.6 & 21 & 16  \\
L694s & $3.6 \pm 5.4$ & $-0.3 \pm 5.9$ & $1.8 \pm 4.0$ & 0.5 & 0.5 & 17 & 17  \\
L810 & $5.6 \pm 14.0$ & $10.7 \pm 11.0$ & $8.6 \pm 8.8$ & 1.0 & 0.3 & 220 & 460  \\

\enddata
\tablenotetext{a}{Line of sight magnetic field, $\mu$G}
\tablenotetext{b}{Mass, $M_\odot$}

\label{results2}
\end{deluxetable}

\clearpage

\clearpage

\end{document}